\documentclass[aps, prl, reprint, superscriptaddress, floatfix, twocolumn]{revtex4-2}
\usepackage[english]{babel}
\usepackage[utf8]{inputenc}
\usepackage[T1]{fontenc}
\usepackage{hyperref, xcolor, graphicx}
\usepackage{amsfonts, amssymb, amsmath, mathtools}

\newcommand\beq{\begin{equation}}
\newcommand\eeq{\end{equation}}
\newcommand\bem{\begin{pmatrix}}
\newcommand\eem{\end{pmatrix}}
\newcommand{\tr}{\mathrm{Tr}}

\newcommand{\mO}{\mathcal{O}}

\newcommand{\la}{\langle}
\newcommand{\ra}{\rangle}

\hypersetup{
    colorlinks=true,
    linkcolor=black,
    citecolor=blue,   
    urlcolor=blue,
}

\begin{document}

\title{Crystalline Spectral Form Factors}

\author{Dmitrii~A.~Trunin}
    \email{dmitrii.trunin@princeton.edu}
    \affiliation{Department of Physics, Princeton University, Princeton, New Jersey 08544, USA}
\author{David~A.~Huse}
    \affiliation{Department of Physics, Princeton University, Princeton, New Jersey 08544, USA}
\date{\today}

\begin{abstract}
    We investigate crystalline-like behavior of the spectral form factor in unitary quantum systems with extremely strong eigenvalue repulsion. Using a low-temperature Coulomb gas as a model of repulsive eigenvalues, we derive the Debye-Waller factor suppressing periodic oscillations of the spectral form factor and estimate the order of its singularities at multiples of the Heisenberg time.  We also reproduce this crystalline-like behavior using perturbed permutation circuits and random matrix ensembles associated with Lax matrices. Our results lay a foundation for future studies of quantum systems that exhibit intermediate level statistics between standard random matrix ensembles and permutation circuits.
\end{abstract}

\maketitle

\textit{\textbf{Introduction}}---Spectral statistics and random matrix theory (RMT) are cornerstones of quantum chaos and quantum statistical mechanics~\cite{Porter,Mehta,Forrester,Haake,Stockmann,Guhr:1997ve,DAlessio:2015}. Most quantum systems fall into one of the four major RMT classes~\footnote{A complete classification of random matrix ensembles reveals additional spectral features~\cite{Altland:1997,Kawabata:2018gjv}.} distinguished by the probability density function of eigenvalue (energy level) spacings~$s$, $P(s) \sim s^\beta$ as $s \to 0$. Quantum chaos is associated with Wigner-Dyson statistics, which corresponds to $\beta = 1, 2$, or $4$ and implies level repulsion~\cite{Bohigas:1984}. On the other hand, integrable systems and localized systems lack level correlations and follow Poisson statistics with $\beta = 0$~\cite{Berry:1977a}. Since the inception of RMT in the framework of nuclear physics~\cite{Weidenmuller:2008vb}, these four universal classes have been observed in a wide range of quantum systems and provided essential tools for quantum many-body dynamics~\cite{Srednicki:1994,Rigol:2008,Sachdev,Nandkishore:2014,Abanin:2018}, quantum information theory~\cite{Collins:2016}, quantum optics~\cite{Carmichael}, quantum transport~\cite{Beenakker:1997}, quantum gravity~\cite{Brezin:1978,Cotler:2016fpe,Saad:2019lba,Cotler:2020ugk,Belin:2023efa,Jafferis:2025vyp}, and many other areas of physics.

However, some quantum systems fall outside the standard RMT classes with $\beta = 0, 1, 2$, or $4$. One class of remarkable examples of nonconventional yet universal level statistics is provided by quantum cellular automata~\cite{Farrelly:2019zds,Arrighi:2019uor,Gopalakrishnan:2018,Gopalakrishnan:2018rfu,Gopalakrishnan:2018wyg,Alba:2019okd,Iaconis:2019hab,Iaconis:2021,Feldmeier:2020xxb,Hillberry:2024xpb} and random permutation circuits~\cite{Bertini:2024tuo,Szasz-Schagrin:2025dil,Bertini:2025vvz}. In a properly chosen basis, discrete-time unitary evolution of these models breaks down into large periodic cycles, so their quasienergy spectrum reduces to sets of roots of unity~\footnote{Possibly shifted by a constant phase.}. Once the length of the largest cycle is comparable to the dimension of the Hilbert space (or dynamics is restricted to one of the invariant subspaces), quasienergy levels become equally spaced on the unit circle. The probability density function of such a ``picket fence'' ensemble is exactly zero in the vicinity of level spacing $s = 0$, so this behavior corresponds to $\beta = \infty$. At the same time, these models are readily engineered, exhibit nontrivial dynamics, and provide a remarkable example of ergodic but not chaotic quantum dynamics (see~\cite{Vikram:2022xiq,Gesteau:2023rrx,Ouseph:2023juq,Camargo:2025zxr} for definitions of quantum ergodic hierarchy). So, it becomes increasingly interesting to understand the emergence of quantum chaos in the entire range $\beta \in (0, \infty)$, and also to find analytically tractable models that show behavior that interpolates between $\beta = \infty$ and standard RMT classes.

\begin{figure}[t!]
    \centering
    \includegraphics[width=\linewidth]{SFF_d=512_beta=500.pdf}
    \caption{The SFF of the circular $\beta$ ensemble, which describes a Coulomb crystal of $d$  repulsive eigenvalues at inverse temperature $\beta$.  We illustrate this for $d=512$, $\beta=500$, averaged over $10^6$ samples. Dashed line shows the damping Debye-Waller factor. Inset: deviation of the SFF from the plateau (enlargement in the same graph).}
    \label{fig:SFF-1}
\end{figure}

In this Letter, we examine three models covering the lightly explored regime between $\beta \sim 1$ and $\beta = \infty$: \hbox{circular} $\beta$ ensemble (Coulomb crystal of eigenvalues), perturbed permutation circuit, and random Lax matrix ensemble. We study correlations between eigenvalues using the Fourier transform of the two-level distribution, also known as the spectral form factor (SFF),
\beq \label{eq:SFF-def}
K(t) = \left\la \left| \tr \, U(t) \right|^2 \right\ra = \sum_{m,n = 0}^{d-1} \left\la e^{i ( E_m - E_n ) t} \right\ra. \eeq
Here, $U(t)=\left[ U(1) \right]^t$ is the evolution operator for integer time $t$, $E_n$ are the corresponding (quasi)energies, $d$ is the dimension of the Hilbert space, and $\la \ldots \ra$ is the ensemble average over statistically similar systems. Since the models that we study exhibit long-range (in energy) crystalline-like correlations of the eigenvalues of $U=U(1)$, we will employ the analogy with Bragg diffraction and treat the SFF as the structure factor of an eigenvalue crystal that is subject to fluctuations~\footnote{Note also an analogy between the eigenvalues of a random matrix ensemble and a system of spinless fermions with long-range repulsive potential~\cite{Forrester}, which also implies oscillations of the SFF~\cite{Krivnov-1,Krivnov-2}.}.  Following this analogy, quasienergies $E_m$ correspond to positions of atoms, and time $t$ corresponds to the momentum change of the scattered wave. 

The SFF is the simplest nontrivial quantity that captures eigenvalue correlations over the entire spectrum. In integrable systems the SFF decreases, in many cases monotonically, from $K(0) = d^2$ to $K(t) \to d$ as $t \to \infty$. In conventionally quantum chaotic systems, it exhibits a telltale ``dip---ramp---plateau'' structure, which reflects level repulsion~\cite{Haake,Stockmann,Guhr:1997ve}. However, none of these standard SFF types show structure at long times $t$ that are large multiples of $d$. The models that we study in this Letter all can exhibit crystalline-like oscillations of the SFF extending to $t\gg d$ due to extremely strong level repulsion (Fig.~\ref{fig:SFF-1}). Similarly to Bragg diffraction from a real crystal, these oscillations are damped due to the weak fluctuations in the level patterns once we move away the exact $\beta = \infty$ limit~\cite{Ashcroft}. We calculate the damping Debye-Waller factor for all mentioned models and confirm it with numerical results. We also extend this calculation to higher-order derivatives in the case of the circular $\beta$ ensemble and show that well-known features of the SFF for the standard cases $\beta = 1,2,4$ are remnants of Bragg peaks.

\textit{\textbf{Coulomb crystal of eigenvalues}}---The circular $\beta$ ensemble extends the Wigner-Dyson level statistics of unitary operators from $\beta = 1, 2, 4$ to an arbitrary real non-negative $\beta$.  The joint probability distribution for the quasienergies $E_n$ of a unitary matrix drawn from the circular $\beta$ ensemble is described by the $d$-particle Coulomb gas confined to a unit circle at temperature $T = 1/\beta$~\footnote{The equilibration of the Coulomb gas is described by the Dyson Brownian motion~\cite{Porter,Mehta,Forrester} and reproduces probability distribution~\eqref{eq:Coulomb-P}. We emphasize that this description does not rely on the specific properties of $\beta = 1, 2, 4$.},
\beq \label{eq:Coulomb-P} \begin{aligned}
P(E_0, \ldots, E_{d-1}) &= Z_{d, \beta}^{-1} \, e^{-\beta H(E_0, \ldots, E_{d-1})}, \\ H(E_0, \ldots, E_{d-1}) &= - \sum_{m < n} \log \left| e^{i E_m} - e^{i E_n} \right|,
\end{aligned} \eeq
where $Z_{d,\beta}$ is the normalizing factor (partition function). 

We emphasize that, in general, the probability distribution of the random matrices that realize the circular $\beta$ ensemble is \textit{not} invariant under basis transformations. Only the standard values $\beta = 1, 2, 4$ correspond to basis-invariant ensembles~\footnote{Physically sound basis transformations are described by an associative division algebra over real numbers. According to the Frobenius theorem, all such algebras are isomorphic to either real, complex, or quaternion numbers. These algebras correspond to the orthogonal ($\beta = 1$), unitary ($\beta = 2$), and symplectic ($\beta = 4$) basis transformations, respectively.}. Nonetheless, level statistics~\eqref{eq:Coulomb-P} with arbitrary $\beta > 0$ does emerge in basis-specific settings, including a class of random-walk-like \hbox{models~\cite{Killip:2004,Killip:2006,CMV}}. This relationship allows us to sample the circular $\beta$ ensemble at low computational cost and study its spectral properties beyond $\beta = 1, 2, 4$. Furthermore, it opens a way for experimental tests and many-body generalizations of our models; see Supplemental Material
and Ref.~\cite{Kolganov:2024nzi} for examples of quantum circuits implementing the circular $\beta$ ensemble.

To introduce the random-walk-like model of the circular $\beta$ ensemble, consider a quantum particle on a discrete interval with coordinates $n = 0, 1, \ldots, l-1$ and an additional internal degree of freedom with basis states~$| \pm \ra$ (thus, $d = 2l$). The discrete-time evolution of the particle consists of alternating shift and scattering steps,
\beq \label{eq:U-CMV}
U = S L S^\dag M~. \eeq
The shift operator $S$ is conditional: it always flips the particle but moves it only if it occupies the $| + \ra$ state,
\beq S | n \ra | - \ra = S | n \ra | + \ra~, \quad S | n \ra | + \ra = |n + 1 \ra | - \ra~. \eeq
The scattering steps are described by $2l \times 2l$ block-diagonal matrices commuting with the position operator,
\beq L = \Theta_1 \oplus \Theta_3 \oplus \ldots \oplus \Theta_{2l-1}~, ~ M = \Theta_0 \oplus \Theta_2 \oplus \ldots \oplus \Theta_{2l-2}~, \eeq
where each $2 \times 2$ block acts on the corresponding fixed-position subspace. 
Blocks $n = 0, \ldots, 2l-2$ are determined by complex numbers $\alpha_n = e^{i \phi_n} \cos \theta_n$, which control the mixing of $| \pm \ra$ states, 
\beq \Theta_n = \bem e^{-i \phi_n} \cos \theta_n & \sin \theta_n \\ \sin \theta_n & - e^{i \phi_n} \cos \theta_n \eem. \eeq
We assume that these coefficients are independently distributed with the following probability density functions:
\beq P(\alpha_n) \propto \left(1 - |\alpha_n|^2 \right)^{(\beta (2l - 1 - n)/2) - 1}. \eeq
The last block is diagonal and depends on a single real parameter uniformly distributed on the interval $(-\pi, \pi)$: $\Theta_{2l-1} = e^{i \phi_{2l-1}} \oplus 1$. This specifies the model: it is proven in~\cite{Killip:2004,Killip:2006} that the eigenvalues of unitaries~\eqref{eq:U-CMV} are distributed according to the circular $\beta$ ensemble for any $d = 2l$ (the model for an odd $d$ is defined similarly).

\textit{\textbf{SFF of the Coulomb gas}}---Before deriving the SFF of the circular $\beta$ ensemble for arbitrary $\beta$, let us first consider the low-temperature limit $\beta \to \infty$. In this limit, repulsion between Coulomb gas particles overcomes thermal fluctuations, but periodic boundary conditions keep particles confined. Hence, the gas condenses into a one-dimensional Coulomb crystal $E_n = (2 \pi n + \phi_{d-1}) / d$, where $\phi_{d-1}$ is the overall phase (center of mass position). In this limit, the SFF becomes purely periodic in $t$ with period $d$: for $t=nd$ for $n$ any integer $K(t)=d^2$.  For all other (integer) values of $t$, $K(t)=0$.
In terms of the random-walk-like model~\eqref{eq:U-CMV}, the limit $\beta \to \infty$ sets all coefficients $\alpha_n$ to zero and turns scattering blocks into swap operators, $\Theta_n \to \sigma_x$. Hence, the evolution of the particle reduces to a cyclic permutation of basis states.

For a large but finite $\beta$, periodic oscillations of the SFF are suppressed at large $t$ by thermal fluctuations~\footnote{See also Refs.~\cite{Forrester,Gorin:2004sfr,Gorin:2024,Desrosiers:2009lye,Majumdar:2009,Forrester:2025} for the discussion of other features of large $\beta$ ensembles.}. Assuming that these fluctuations are relatively small~\footnote{Expansion near infinite-temperature eigenvalues is self-consistent if the fluctuation amplitude is smaller than average level spacing, i.e., $\Delta x_k^2 \approx 2 \log d / \beta d^2 < (2 \pi / d)^2$. Hence, it is justified for $\beta > 1$ and any numerically feasible system size ($d < 10^8$).}, we expand the Hamiltonian~\eqref{eq:Coulomb-P} near the saddle point, $E_n = E_n^{(\beta=\infty)} + x_n$ with $x_n \ll 2\pi/d$, neglect higher-order nonlinear terms, and switch to the momentum picture,
\beq \label{eq:Coulomb-Gauss}
H \approx \frac{1}{2} \sum_{p=1}^{d-1} p (d - p) \tilde{x}_p \tilde{x}_{-p}~, \eeq
where $\tilde{x}_p = \frac{1}{\sqrt{d}} \sum_{k=0}^{d-1} e^{-2 \pi i k p / d} x_k$. Note that the center-of-mass degree of freedom ($p = 0$ mode) is static, so we integrate it out. In this approximation, we explicitly estimate the ensemble average in the SFF~\eqref{eq:SFF-def},
\beq \label{eq:SFF-Coulomb} \begin{aligned}
    K(t) &= \sum_{j,k} e^{2 \pi i (k - j) t / d} \left\langle e^{i t (x_k - x_j)} \right\rangle \\
    &\approx d + d \sum_{k=1}^{d-1} e^{2 \pi i k t / d} \left| C d \sin(\pi k / d) \right|^{- 4 t^2 / \beta d^2}.
\end{aligned} \eeq
We assume $d \gg 1$ to approximate the sums involving two-point correlation functions. The numerical coefficient $C \approx 3.6$. Similarly to the zero-temperature SFF, ``scattering amplitudes'' constructively interfere near multiples of the Heisenberg time $t_H = d$. Hence, the finite-temperature SFF oscillates with period $t_H$. The effect of thermal vibrations is contained in the Debye-Waller factor $e^{-2W}$,
\beq K(t_H \tau) \approx d + e^{-2 W} d^2, \eeq
which suppresses peaks of the SFF ($\tau = t / t_H = 1, 2, \ldots$),
\beq \label{eq:Debye-Waller}
e^{-2W} \sim \begin{cases} d^{- 4 \tau^2 / \beta}, \quad & 4 \tau^2 / \beta \ll 1~, \\ \log(d) / d~, \quad & 4 \tau^2 / \beta = 1~, \\ (\beta / \tau^2 d) (C \pi)^{- 4 \tau^2 / \beta}, \quad & 4 \tau^2 / \beta \gg 1~. \end{cases} \eeq
So, the peaks become negligible, and the SFF reaches the plateau only at times $t \gg t_* \approx t_H \sqrt{\beta / 4}$. This new timescale is much larger than the Heisenberg time if $\beta \gg 1$.

\begin{figure}[t]
    \centering
    \includegraphics[width=\linewidth]{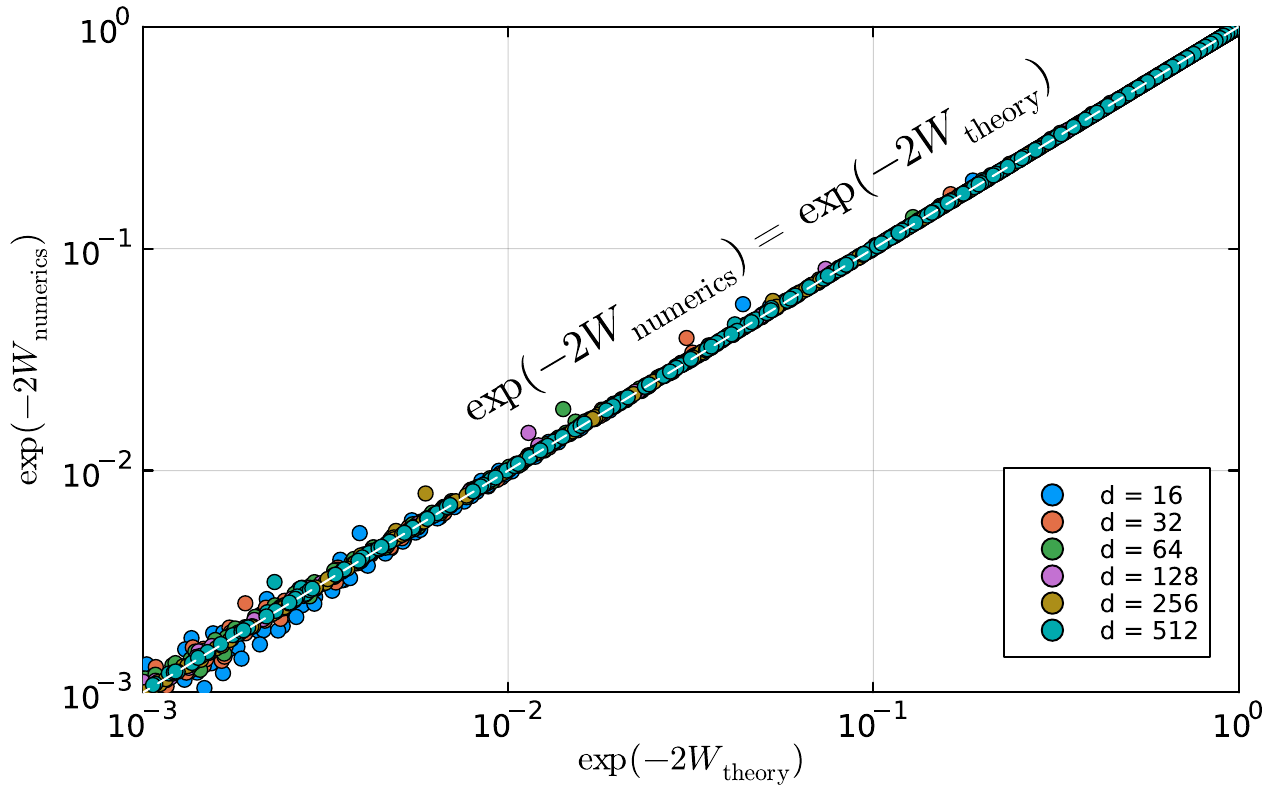}
    \caption{Comparison of numerical and theoretical Debye-Waller factors of the circular $\beta$ ensemble. Points for inverse temperatures $1 < \beta < 20000$ are superposed.}
    \label{fig:SFF-2}
\end{figure}

The numerical calculations based on the random-walk-like model~\eqref{eq:U-CMV} agree well with theoretical estimates, e.g., see Figs.~\ref{fig:SFF-1} and~\ref{fig:SFF-2}. In particular, we confirmed that full expression~\eqref{eq:SFF-Coulomb} and Debye-Waller factor~\eqref{eq:Debye-Waller} approximate the exact SFF with an error less than several percent when $\beta > 10$ and $e^{-2W} > 0.01$.

Furthermore, the Gaussian approximation captures the order of the SFF singularities at multiples of the Heisenberg time. Differentiating~\eqref{eq:SFF-Coulomb} by $\tau = t/t_H$ in the limit $d \to \infty$, we find that the $n$th derivative $K^{(n)}(\tau)$ at $\tau = 1, 2, \ldots$ is divergent for $n \ge \gamma$ (merely discontinuous if $n = \gamma$ and $n$ is odd), where
\beq \label{eq:SFF-disc}
\gamma = 4 \tau^2 / \beta - 1. \eeq
In particular, this formula recovers the exact results~\cite{Mehta,Forrester} for $\tau = 1$ ($\beta = 1, 2, 4$) and $\tau = 2$ ($\beta = 4$), and shows that even for these standard RMT cases, the spectrum can be viewed as a strongly fluctuating crystal, with a singularity remaining at its first (and second for $\beta=4$) Bragg peak.  However, we emphasize that Gaussian approximation~\eqref{eq:SFF-Coulomb} works only for relatively small times $\tau \ll \beta$ (see Supplemental Material
for details). So, in general, the estimate~\eqref{eq:SFF-disc} is guaranteed to be valid only for $\tau \ll \beta$, even though properties of a given $\beta$ can significantly extend this region~\footnote{For a rational $\beta / 2 = p / q$ with $p$ and~$q$ mutually prime, the asymptotic expansion of the density-density correlation function of the circular $\beta$ ensemble, Eq.~(13.226) in~\cite{Forrester}, implies that the SFF has no singularities for $\tau > p$, whereas Eq.~\eqref{eq:SFF-disc} is valid for $\tau \le p$.}. This limitation explains the discrepancy between Eq.~\eqref{eq:SFF-disc} and exact $\beta = 1, 2, 4$ results at larger times, where the SFF in those specific cases has no singularities. Nevertheless, it leaves ample room for conspicuous crystalline-like behavior of the SFF for any $\beta \gg 1$.

\begin{figure}[t]
    \centering
    \includegraphics[width=\linewidth]{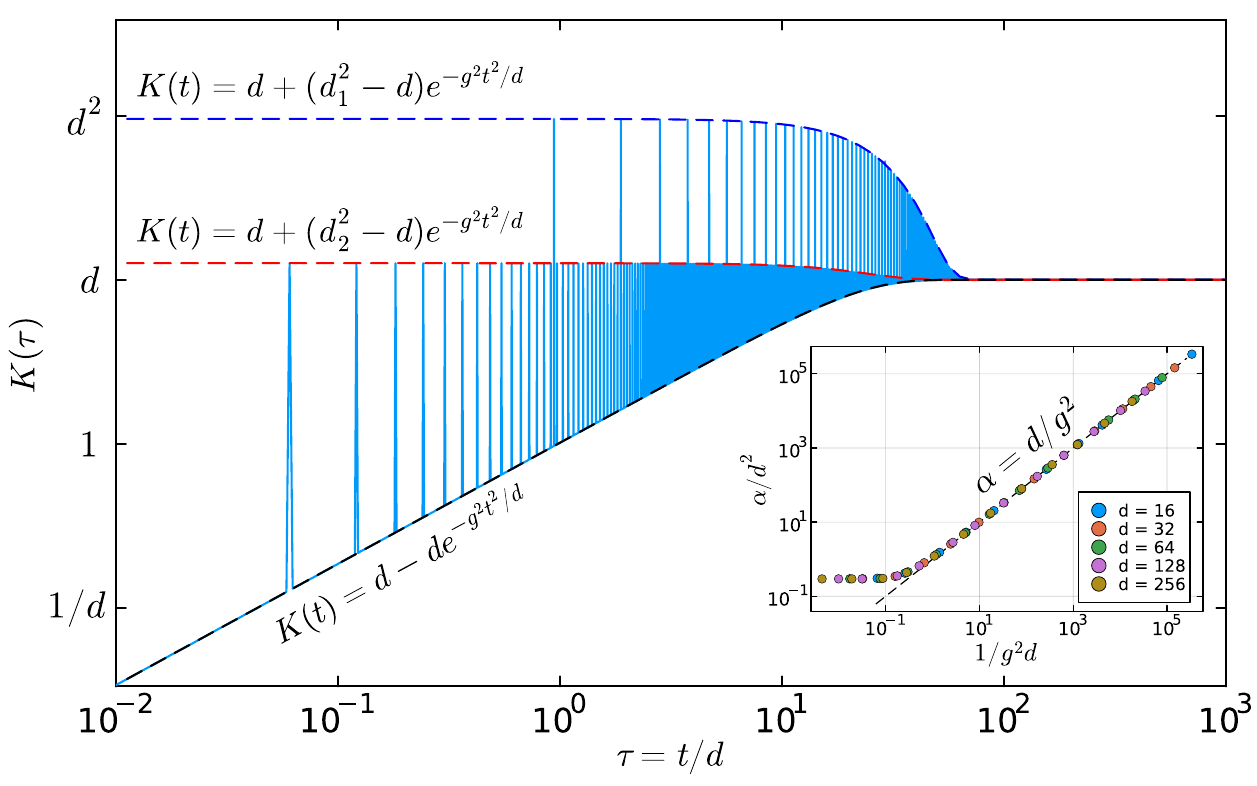}
    \caption{The SFF of toy model~\eqref{eq:Interpolation} interpolating between $\beta = 2$ and $\beta = \infty$. We set $d = 512$, $g = 0.002$, and average over $10^5$ samples. The base permutation $S$ contains two cycles of length $d_1 = 481$ and $d_2 = 31$. Inset: strength of the pinning potential $V(x) = \alpha x^2 / 2$ determined by fitting the SFF with Eq.~\eqref{eq:SFF-Interpolation}. The permutation used in the inset contains a single cycle.}
    \label{fig:SFF-3}
\end{figure}

\textit{\textbf{Perturbed permutation circuit}}---There are other models interpolating between the standard RMT universality classes. Besides empirical Brody~\cite{Brody:1973,Brody:1981} and Berry-Robnik~\cite{Berry:1984a} distributions, explicit examples of intermediate level statistics include \hbox{banded matrices~\cite{Moshe:1994gc,Mirlin:1996,Bogomolny:2010,Evers:2008},} 
Pandey-Mehta~\cite{Pandey:1982br,Lenz:1990,Lenz:1991,Dupuis:1991}, Rosenzweig-Porter~\cite{Rosenzweig:1960,Kravtsov:2015}, and other toy models~\cite{Izrailev:1990elx,Prosen:1993,Prosen:1994,Prosen:1997,Prosen:2002,Prosen:2018,Chan:2018dzt,Friedman:2019gyi,Sierant:2020}. 
However, these models are restricted to the interval between $\beta = 0$ and $\beta = 2$.  Hence, they cannot produce the crystalline-like behavior of the SFF that is present at large $\beta$.  We suggest a new simple model, which interpolates between $\beta=2$ and $\beta = \infty$,
\beq \label{eq:Interpolation}
U = \exp(-i g H) \, S~. \eeq
Here, $H$ is a random $d \times d$ Hermitian matrix drawn from the Gaussian unitary ensemble with probability density $\propto \exp\!\left[- (d / 2) \tr H H^\dag \right]$, $S$ is a classical permutation of basis vectors supplemented with independent and uniformly distributed random phase rotations, $S | k \ra = e^{i \phi_k} | \pi(k) \ra$, and $0 \le g \le \pi/2$ is a real parameter controlling the strength of this Gaussian unitary ensemble perturbation to the permutation. For simplicity, we consider global matrices $H$ and $S$; however, model~\eqref{eq:Interpolation} is straightforwardly reproduced by a local shallow circuit~\footnote{We discuss a straightforward local generalization of model~\eqref{eq:Interpolation} in the Supplemental Material
. Besides, local permutation circuits with exponentially large cycles can be constructed using primitive polynomials over the binary field, e.g., see~\cite{Ippoliti:2025oro,Golomb}.}. We also assume that the permutation~$\pi(k)$ is fixed but drawn at random. In this case, the typical length of its largest cycle is comparable to $d$~\cite{Ford:2022}, and the Heisenberg time $t_H = \mO(d)$.

Similarly to the circular $\beta$ ensemble, we first consider the limit $g \to 0$. In this limit, eigenvalues of unitary~\eqref{eq:Interpolation} weakly fluctuate near the eigenvalues of the permutation matrix $S$, so we can use perturbation theory to evaluate the SFF. In general, permutation $\pi(k)$ contains $N$ cycles of length $d_n$, and each cycle produces a series of equally spaced SFF peaks,
\beq \label{eq:SFF-Interpolation}
K(t) \approx d + e^{-t^2g^2 /d} \left[ \left( \sum_{n=1}^N d_n \delta_{t \bmod d_n, 0} \right)^2 - d\right]. \eeq
In other words, the model describes $N$ independent crystals of $d_n$ noninteracting particles (eigenvalues) pinned to the corresponding lattices by a harmonic potential $V(x) = \alpha x^2 / 2$ with $\alpha = d/g^2$, with temperature one.  

For small but finite $g \ll 1 / \sqrt{d}$, eigenvalue fluctuations are smaller than level spacings, so the structure of the SFF~\eqref{eq:SFF-Interpolation} and the damping Debye-Waller factors are qualitatively preserved, see Fig.~\ref{fig:SFF-3}. In this limit, the SFF reaches the plateau for $t \gg t_* \sim \sqrt{d \log d} / g \gg t_H$. In the opposite limit $d \gg 1/g^2$, repulsion between eigenvalues recovers the circular unitary ensemble, so the SFF approaches the standard ``dip-ramp-plateau'' form after the Thouless time $t_\mathrm{Th} \sim \sqrt{d} / g \ll t_H$.  Thus as $dg^2$ is varied, this system interpolates between a perfect crystal that matches $\beta\rightarrow\infty$ and the $\beta=2$ circular unitary ensemble, but in between these two limits its SFF is different from the intermediate circular $\beta$ ensembles.

\begin{figure}[t]
    \centering
    \includegraphics[width=\linewidth]{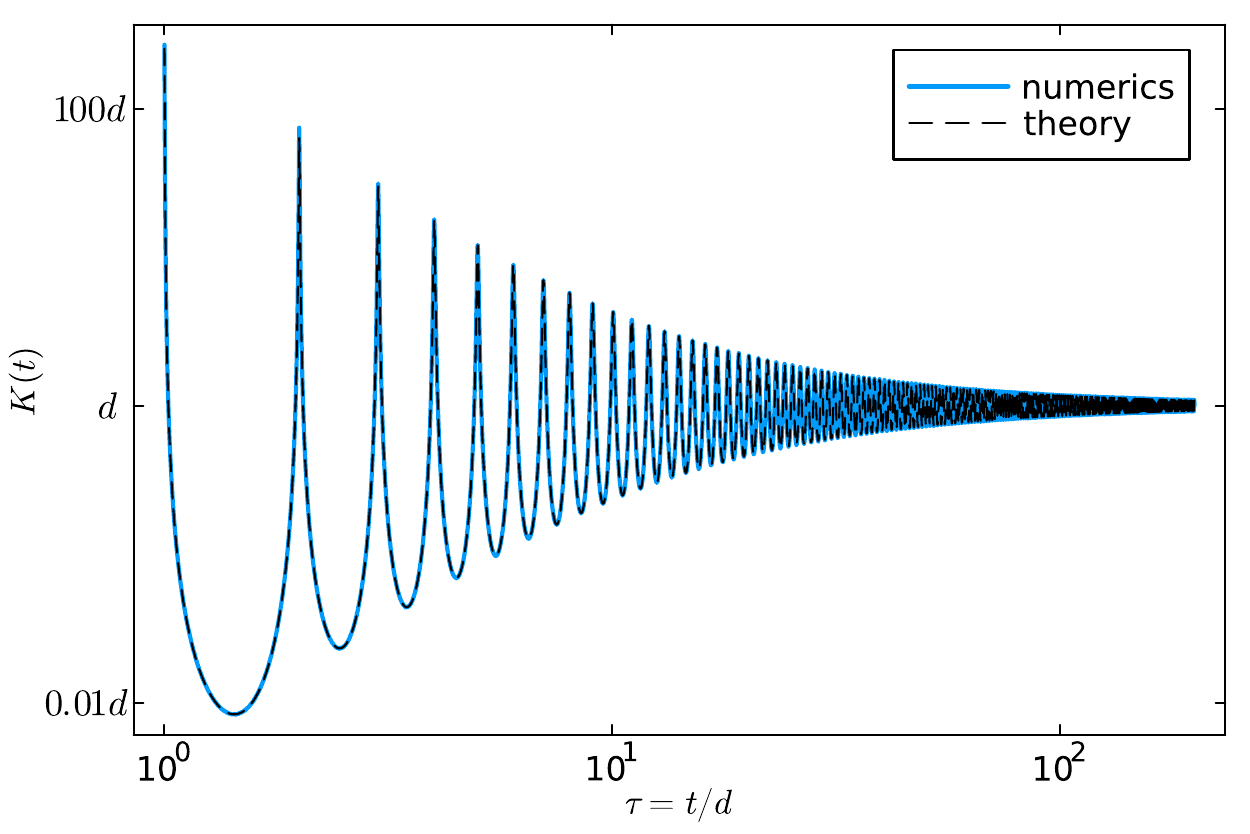}
    \caption{SFF of the random ensemble of Lax matrices~\eqref{eq:Lax}. We set $d = 512$, $g = 0.98$, and average over $10^5$ samples.}
    \label{fig:SFF-4}
\end{figure}

\textit{\textbf{Random Lax matrix ensemble}}---The last example of the crystalline SFF is given by the ensemble of random Lax matrices $U$ of the Ruijsenaars-Schneider model~\cite{Bogomolny:2009,Bogomolny:2011}, which also emerges as quantization of interval-exchange maps~\cite{Bogomolny:2004,Giraud:2004},
\beq \label{eq:Lax}
U_{mn} = \frac{1}{d} e^{i p_m} \frac{1 - e^{2 \pi i g}}{1 - e^{2 \pi i (m-n+g)/d}}~, \eeq
where $p_m$ are independent random variables uniformly distributed between $-\pi$ and $\pi$, $d$ is the dimension of the Hilbert space, and parameter $0 < g < 1$ controls level repulsion. In the limit $g \to 0$, unitary~\eqref{eq:Lax} becomes diagonal and has Poisson level statistics. In the limit $g \to 1$, it turns into a cyclic shift operator supplemented with random phase rotations: $U_{mn} = e^{i p_m} \delta_{m+1,n}$, so has a zero-temperature eigenvalue crystal.  Thus these two limits correspond to $\beta=0$ and $\beta=\infty$, respectively.  For intermediate values of $g$, the spectral statistics of $U$ is described by a gas of hard rods~\cite{Bogomolny:2011},
\beq \label{eq:Lax-P}
P(E_0, \ldots, E_{d-1}) \propto \prod_{j=0}^{d-1} \theta\!\left( E_{j+1} - E_j - \frac{2 \pi}{d} g \right)~, \eeq
where 
$g$ is the scaled length of the rods, and $E_j$ are the positions of the rod centers.  Thus this system provides an interpolation between $\beta=0$ and $\beta=\infty$; again, between these two limits this system differs from the circular $\beta$ ensembles. 
For large $d$, the SFF behaves as
\beq \label{eq:SFF-Lax}
\frac{K(t=\tau d)}{d} \approx 1 + 2 \mathrm{Re} \frac{1}{e^{2 \pi i g \tau} \left[ 1 + 2 \pi i (1-g) \tau \right] - 1}~. \eeq
This result is exact in the limit $d \to \infty$~\cite{Bogomolny:2011} and agrees with numerical results for $d \gg 1$, e.g., see Fig.~\ref{fig:SFF-4}. Similarly to the SFF of the Coulomb gas~\eqref{eq:SFF-Coulomb}, this SFF oscillates in an earlier time regime with period $t_H \approx d$ when $1 - g \ll 1$. Not so similarly, $K(t) / d$ and all of its time derivatives remain finite for $0 < g < 1$ and any $\tau = t / d>0$ in the limit $d \to \infty$.  For $0<t\ll d/(1-g)$ the peaks in $K(t)/d$ are of height $\sim [d/((1-g)t)]^2\gg 1$ and thus strong.  For later times, $K(t)/d$ continues to weakly oscillate, but now with a longer period $d/g$ and amplitude $\sim d/((1-g)t)\ll 1$.

\textit{\textbf{Conclusion}}---We have presented three models of ``crystalline'' level statistics associated with oscillations of the SFF. For all three models, we have analytically derived expressions for the Debye-Waller factor, which suppresses oscillations of the SFF, and the new timescale $t_* \gg t_H$, at which the SFF approaches the plateau. This timescale is a conspicuous feature of intermediate level statistics and thus provides the easiest practical way to quantify long-range correlations between eigenvalues in the range between $\beta \sim 1$ and $\beta = \infty$. The key ingredient ensuring the crystalline-like behavior in all considered models is strong level repulsion. Our examples indicate that this behavior occurs in some broader class of chaotic quantum systems. We also emphasize that this class is not restricted to unitary discrete-time evolution, although the examples we gave are of this type. In particular, the necessary requirement for crystalline level statistics is satisfied by the Gaussian $\beta$ ensemble at large~$\beta$~\cite{Dumitriu:2002}, the weakly disordered one-dimensional Anderson localization model~\cite{Torres-Herrera:2019}, an ensemble of random Lax matrices of Calogero-Moser model~\cite{Bogomolny:2009,Bogomolny:2011}, and inhomogeneous discrete-time quantum walks~\cite{Cantero:2010,Linden:2009,Ambarish:2017,Gong:2022,Kempe:2003,Venegas-Andraca:2012zkr,Aharonov:1993gdz,Ambainis:2001,Ryan:2005}. It can also be extended to higher dimensions following~\cite{Massaro:2025}.

We expect that our work paves a way for further investigations in various directions.  First, crystalline-like eigenvalue statistics implies that dephasing in the eigenstate basis is delayed until time $t_* \gg t_H$, so it suggests a careful study of the behavior of correlation functions, entanglement, complexity, and other essential tools of quantum dynamics, for our examples between $t_H$ and $t_*$. We emphasize  that systems with $1 \ll \beta \ll \infty$ are ergodic, but not necessarily chaotic (or even mixing), so they provide a perfect testing ground for measures of quantum chaoticity, which are expected to distinguish between different levels of quantum ergodic hierarchy~\cite{Vikram:2022xiq,Gesteau:2023rrx,Ouseph:2023juq,Camargo:2025zxr}. In particular, the crystalline structure of the SFF serves as the simplest and most evident signature of violated (or significantly delayed) quantum mixing.

Second, it would be interesting to find and investigate many-body systems with crystalline-like level statistics intermediate between $\beta \sim 1$ and $\beta = \infty$. A straightforward generalization of the discrete-walk-like model~\eqref{eq:U-CMV} to the case of free noninteracting fermions, which we describe in the Supplemental Material
, resembles other toy models of the many-body SFF~\cite{Liao:2020lac,Winer:2020mdc,Flynn:2024thf,Ikeda:2024mgb} and serves as a good starting point for this endeavor. Besides, models of intermediate level statistics between $\beta = 0$ and $\beta \sim 1$ play an important role for the studies of many-body-localization transitions~\cite{Oganesyan:2006,Atas:2012,Buijsman:2019,Prakash:2020psj}, so it would be interesting to find an analog of this transition between $\beta \sim 1$ and $\beta = \infty$.

Third, we expect that oscillations of the SFF in low-temperature critical models in the vicinity of the critical point~\cite{Nivedita:2020njr,Dyer:2016pou,Benjamin:2018kre} are suppressed at large evolution times similarly to our models; it would be interesting to check this prediction.

Finally, recent experimental advancements~\cite{Dong:2024yaf} make it tempting to experimentally confirm the crystalline-like behavior of the SFF in our models.

\textit{\textbf{Acknowledgements}}---The authors thank Sarang Gopalakrishnan, Amit Vikram, and Peter Forrester for helpful discussions.  D. A. H. was supported in part by NSF QLCI grant OMA-2120757.

\bibliography{letter}

\end{document}


\title{Supplemental Material for ``Crystalline Spectral Form Factors''}

\maketitle

Here, we report some useful information supplementing the main text. In particular:
\begin{itemize}
    \item In Sec.~\ref{sec:Coulomb}, we introduce quantum circuit models of the one-dimensional Coulomb gas (circular $\beta$-ensemble);
    \item In Sec.~\ref{sec:Gaussian-limit}, we discuss the limitations of the Gaussian approximation for the SFF of the circular $\beta$-ensemble;
    \item in Sec.~\ref{sec:Interpolating-SFF}, we derive the Debye-Waller factor of the toy interpolating model $U = e^{-i g H} S$;
    \item in Sec.~\ref{sec:Circuit}, we describe a local shallow circuit that reproduces the toy interpolating model $U = e^{-i g H} S$.
\end{itemize}

\section{Quantum circuits for the circular beta ensemble}
\label{sec:Coulomb}

\begin{figure}[t]
    \centering
    \includegraphics[width=0.5\linewidth]{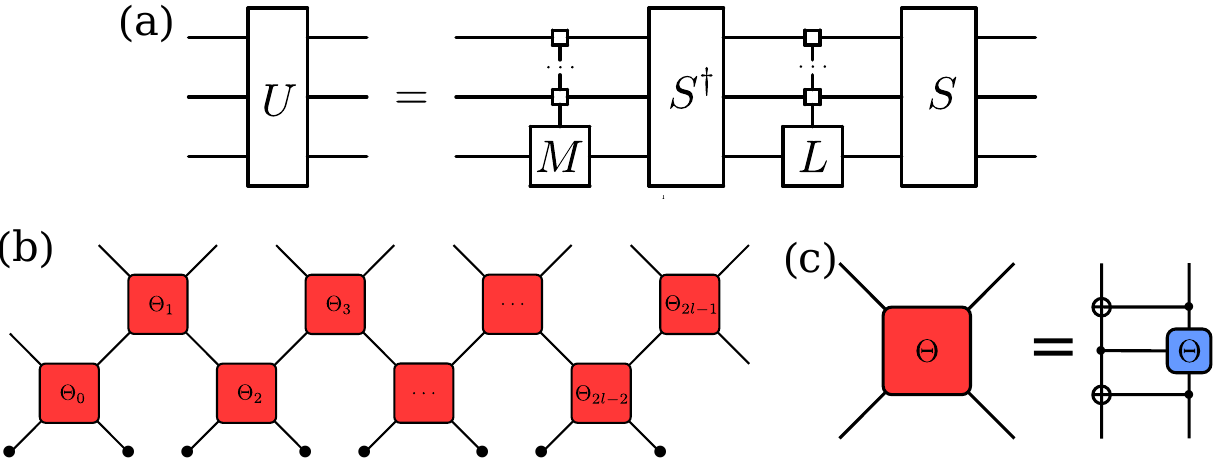}
    \caption{Quantum circuits implementing the random-walk-like model of the circular $\beta$-ensemble. (a) One-particle model on $N$ qubits ($2^N \ge d$). (b) Many-body model on $d$ qubits, which describes non-interacting fermions and reproduces the original model in the one-particle sector. (c) Structure of the two-qubit gates comprising the brick-work circuit (b).}
    \label{fig:circuits}
\end{figure}

There are two simple ways to implement the random-walk-like model $U = S L S^\dag M$ using a quantum circuit. In the first approach, we simply identify the basis of the hopping particle with computational basis states of an $N$-qubit quantum circuit:
\beq | n \ra | \pm \ra \to |2n + (1 \pm 1) / 2 \ra = |2^{N-1} b_{N-1} + \ldots + 2 b_1 + b_0 \ra = \bigotimes_{k=0}^{N-1} |b_k \ra, \eeq
where $b_k = 0, 1$. In this basis, matrix $S$ merely represents a cyclic adding operation, $S |n \ra = |n + 1\ra$, and block-diagonal matrices $L$ and $M$ are nothing but quantum multiplexers, i.e., multi-qubit controlled operations. In other words, they conditionally apply the gate $\Theta_{2k + \lambda}$ to the target $0$-th qubit, where $k = 2^{N-2} b_{N-1} + \ldots + 2 b_2 + b_1$ is determined by the state of control qubits $| b_{N-1} \ra \otimes \cdots \otimes | b_1 \ra$ and $\lambda = 0, 1$ for operators $M, L$, respectively. Note that in general, $2^N \ge d$, where $d = 2l$ is the dimension of the original Hilbert space of the hopping particle; if equality does not hold, we simply set $\Theta_k = \mathbb{I}$ for $k > d$. This circuit is depicted on Fig.~\ref{fig:circuits}(a).

The second approach that implements $U$ in quantum circuits promotes the one-particle random walk to a many-particle one. Namely, we construct the following two-layer brick-work circuit acting on $2l$ qubits (we identify qubits $2l$ and $0$):
\beq \mathcal{U} = \prod_{k=0}^{l-1} \tilde{\Theta}_{2k+1}(2k+1, 2k+2) \prod_{k=0}^{l-1} \tilde{\Theta}_{2k}(2k, 2k+1).  \eeq
In this notation, two-qubit gates $\tilde{\Theta}_k(m, n)$, which act on qubits $m$ and $n$, are constructed from the same single-qubit blocks $\Theta_k$ as matrices $L$ and $M$:
\beq \tilde{\Theta}_k(m, n) = \mathrm{CNOT}(n \to m) \Theta_k(m \to n) \mathrm{CNOT}(n \to m) = \bem 1 & 0 & 0 & 0 \\ 0 & e^{-i \phi_k} \cos \theta_k & \sin \theta_k & 0 \\ 0 & \sin \theta_k & - e^{i \phi_k} \cos \theta_k & 0 \\ 0 & 0 & 0 & 1 \eem, \eeq
where $O(c \to t)$ denotes a controlled-$O$ gate with controlled qubit $c$ and target qubit $t$. Circuit $\mathcal{U}$ clearly preserves the number of particles $Q = \sum_k b_k$ and reproduces the one-particle hopping problem in the subspace $Q = 1$. This circuit is depicted on Fig.~\ref{fig:circuits}(b).

\section{Limits of the Gaussian approximation}
\label{sec:Gaussian-limit}

In this section, we explain why Gaussian approximation for the SFF of the circular $\beta$-ensemble fails at times $t/d \gg \beta$. To that end, we again expand the Hamiltonian of the circular $\beta$-ensemble near the zero-temperature equilibrium positions, $E_n = E_n^{\beta = \infty} + x_n$, but keep the first non-linear term:
\beq \label{eq:H-appr}
H = \frac{1}{2} d^2 \sum_{p=1}^{d-1} S_2\!\left(\frac{p}{d}\right) \tilde{x}_p \tilde{x}_{-p} + i d^{5/2} \sum_{p,q = 1}^{d-1} S_3\!\left(\frac{p+q}{d}\right) \tilde{x}_p \tilde{x}_q \tilde{x}_{-p-q} + \ldots, \eeq
where we introduced the following short notations:
\beq S_2(x) = x (1 - x), \qquad
    S_3(x) = \frac{1}{d^3} \sum_{k=1}^d \frac{\cos(\pi k/d)}{4 \sin^3(\pi k / d)} \sin(2 \pi k x). \eeq
Then we rewrite the SFF in terms of Fourier harmonics $\tilde{x}_p$:
\beq K(t) = \sum_{j,k} e^{2 \pi i (j-k)/d} \left\langle e^{i t (x_j - x_k)} \right\rangle = \sum_{j,k} e^{2 \pi i (j-k)/d} \left\langle e^{i t \sum_p S_{j,k}(p) \tilde{x}_p} \right\rangle, \qquad S_{j,k}(p) = \frac{1}{\sqrt{d}} \left( e^{2 \pi i j p / d} - e^{2 \pi i k p / d} \right). \eeq
Thus, we need to estimate the expectation value $V = \left\la \exp\!\left[i t \sum_p S_{j,k}(0) \tilde{x}_p \right] \right\ra$ in the theory~\eqref{eq:H-appr}. We know that $S_2, S_3 \sim 1$ and that on average, the deviation of modes $ \la \tilde{x}_p^2 \ra \sim \log(d) / \beta d^2$, so higher-order terms of~\eqref{eq:H-appr} should be suppressed in the limit $d \to \infty$. Hence, we can perturbatively estimate the expectation value $V$. In the leading order, it is approximated by the expectation value in a purely Gaussian theory:
\beq V \approx \left\la e^{i t \sum_p S_{j,k}(p) \tilde{x}_p} \right\ra_2 
= \sum_{n=0}^\infty \frac{(i t)^{2n}}{(2n)!} (2n-1)!! \left[ \sum_{p,q} S_{j,k}(p) S_{j,k}(q) \la \tilde{x}_p \tilde{x}_q \ra_2 \right]^n = \exp\!\left[ - \frac{t^2}{2} \sum_p \frac{ S_{j,k}(p) S_{j,k}(-p) }{\beta S_2(p/d)} \right], \eeq
and we know that
\beq \sum_p \frac{ S_{j,k}(p) S_{j,k}(-p) }{\beta S_2(p/d)} = \frac{8}{\beta d^2} \sum_p \frac{1}{p} \sin^2\left( \frac{\pi (j-k) p}{d} \right) \approx \frac{4}{\beta d^2} \log\left| Cd \sin \frac{\pi(j-k)}{d} \right|, \eeq
which determines a natural time scale of the Gaussian theory $t \sim d \sqrt{\beta}$. The leading nonlinear correction to $V$ is described by a diagram with a single cubic vertex:
\beq \begin{aligned}
    \Delta_3 V &\approx \left\la i d^{5/2} \sum_{p,q = 1}^{d-1} S_3\!\left(\frac{p+q}{d}\right) \tilde{x}_p \tilde{x}_q \tilde{x}_{-p-q} e^{i t \sum_p S_{j,k}(p) \tilde{x}_p} \right\ra_2 = \\
    &= \left\la i d^{5/2} \sum_{p,q} S_3\!\left(\frac{p+q}{d}\right) \tilde{x}_p \tilde{x}_q \tilde{x}_{-p-q}  \sum_{n=0}^\infty \frac{(it)^{2n+1}}{(2n+1)!} \left( \sum_p S_{j,k}(p) \tilde{x}_p \right)^{2n+1} \right\ra_2 \\
    &= t^3 d^{5/2} \sum_{p,q,s,t,u} S_3\!\left(\frac{p+q}{d}\right) S_{j,k}(s) S_{j,k}(t) S_{j,k}(u) \left\la \tilde{x}_p \tilde{x}_s \right\ra_2 \left\la \tilde{x}_q \tilde{x}_t \right\ra_2 \left\la \tilde{x}_{-p-q} \tilde{x}_u \right\ra_2 \times \\ &\qquad\qquad\qquad\times \sum_{n=1}^\infty \frac{(it)^{2n-2}}{(2n-2)!} (2n-3)!! \left[ \sum_{p,q} S_{j,k}(p) S_{j,k}(q) \la \tilde{x}_p \tilde{x}_q \ra_2 \right]^{n-1} \\
    &= \frac{t^3 d^{5/2}}{\beta^3} \sum_{p,q} S_3\!\left(\frac{p+q}{d}\right) \frac{S_{j,k}(-p) S_{j,k}(-q) S_{j,k}(p+q)}{p(d-p)q(d-q)(p+q)(d-p-q)} \exp\!\left[ - \frac{t^2}{2} \sum_p \frac{ S_{j,k}(p) S_{j,k}(-p) }{\beta S_2(p/d)} \right] \\
    &\approx \left(\frac{t}{\beta d}\right)^3 \int_0^1 dx \int_0^1 dy \frac{S_3(x+y) S_{j,k}(-x) S_{j,k}(-y) S_{j,k}(x+y)}{x (1-x) y(1-y) (x+y) (1-x-y)} \exp\!\left[ -\frac{t^2}{\beta d^2} \int_0^1 dx \frac{S_{j,k}(x) S_{j,k}(-x)}{S_2(x)} \right], 
\end{aligned} \eeq
where we took into account that correlation functions of odd number of entries are zero and $S_2(0) = S_3(0) = 0$. This expression implies two natural time scales. The first one is the time scale of the Gaussian theory, $t \sim d\sqrt{\beta}$, which determines the exponential decay of this correction to the SFF. The second time scale, $t \sim \beta d$, comes from the prefactor; this time scale is induced by loop corrections, which add extra fields $\tilde{x}$ in the correlators and pull down powers of $1/\beta$, but do not bring with them extra powers of $1/d$. Higher-order nonlinearities and numbers of loops follow the same pattern and become noticeable only for large times $t \gg \beta d$. Thus, Gaussian approximation works well enough for smaller times (at least as long as we are concerned with the SFF).

\section{Spectral Form Factor of the toy Interpolating Model}
\label{sec:Interpolating-SFF}

For weak perturbations above the permutation circuit limit, the SFF of the toy interpolating model $U = e^{-i g H} S$ exhibits a crystalline-like behavior with damped periodic oscillations. In this section, we estimate the damping factor of these oscillations. We work in the limit $g \ll 1$ and $d \gg 1$. We also assume that matrix $H$ is drawn from the Gaussian Unitary Ensemble with partition function $Z = \int dH \, \exp\!\left( -\frac{d}{2} \, \tr H^2 \right)$. With such a normalization, eigenvalues of $H$ fall into the interval $(-2, 2)$. 

For simplicity, we first assume that the permutation of basis vectors contains a single cycle of maximal length~$d$, so $S^d = e^{i \phi} \hat{I}$. In the unperturbed model $U = S$, all eigenvalues are fixed, $E_n = (2 \pi n + \phi)/ d$, $n = 0,1,\ldots,d-1$, and eigenvectors are straightforwardly constructed by following the classical cycle:
\beq | E_n \rangle = \frac{1}{\sqrt{d}} \sum_{k=0}^{d-1} e^{-2 \pi i k n / d} \left| \pi^k(0) \right\ra = \frac{1}{\sqrt{d}} \sum_{k=0}^{d-1} e^{-2 \pi i n \tilde{\pi}(k) / d} | k \ra, \eeq
where $\pi^k(0)$ denotes the $k$-th element of the cycle seeded by $0$ and $\tilde{\pi}(k)$ denotes the discrete logarithm, i.e., the distance between elements $0$ and $k$ along the cycle, $\pi^{\tilde{\pi}(k)}(0) = k$. The leading correction to eigenvalues, $E_n \to E_n + x_n$, is given by the perturbation theory (since $x_n \ll 1$):
\beq e^{i E_n + i x_n} - e^{i E_n} \approx \langle E_n | \left( e^{-i g H} - 1 \right) S | E_n \rangle = e^{i E_n} \langle E_n | \left( e^{-i g H} - 1 \right) | E_n \rangle, \eeq
hence,
\beq x_n \approx - \frac{g}{d} \sum_{a,b = 0}^{d - 1} H_{ab} \, e^{2 \pi i n \left[ \tilde{\pi}(a) - \tilde{\pi}(b) \right] / d}. \eeq
Taking into account that matrices $H_{a,b}$ are distributed according to the Gaussian unitary ensemble, we can estimate the approximate two-point correlation function of eigenvalues in the limit $d \to \infty$:
\beq \label{eq:two-point}
\begin{aligned}
    \langle x_m x_n \rangle &\approx \frac{g^2}{d^2} \sum_{a,b,j,k} e^{2 \pi i m \left[ \tilde{\pi}(a) - \tilde{\pi}(b) \right] / d} e^{2 \pi i n \left[ \tilde{\pi}(j) - \tilde{\pi}(k) \right] / d} \langle H_{ab} H_{jk} \rangle \\ 
    &\approx \frac{g^2}{d^3} \sum_{a,b,j,k} e^{2 \pi i m \left[ \tilde{\pi}(a) - \tilde{\pi}(b) \right] / d} e^{2 \pi i n \left[ \tilde{\pi}(j) - \tilde{\pi}(k) \right] / d} \delta_{ak} \delta_{bj} \\
    &\approx \frac{g^2}{d^3} \sum_{a,b} e^{2 \pi i (m - n) (a - b)} \approx \frac{g^2}{d} \delta_{m,n},
\end{aligned} \eeq
where we used that $k \to \tilde{\pi}(k)$ is a bijection. The momentum-space correlation function is equally flat:
\beq \la \tilde{x}_p \tilde{x}_q \ra = \frac{1}{d} \sum_{m,n} e^{-2 \pi i (p m + q n) / d} \la x_m x_n \ra = \frac{g^2}{d} \delta_{p, -q}, \eeq
which indicates a flat spectrum of quasiparticles in the eigenvalue crystal.

\begin{figure}[t]
    \centering
    \includegraphics[width=\linewidth]{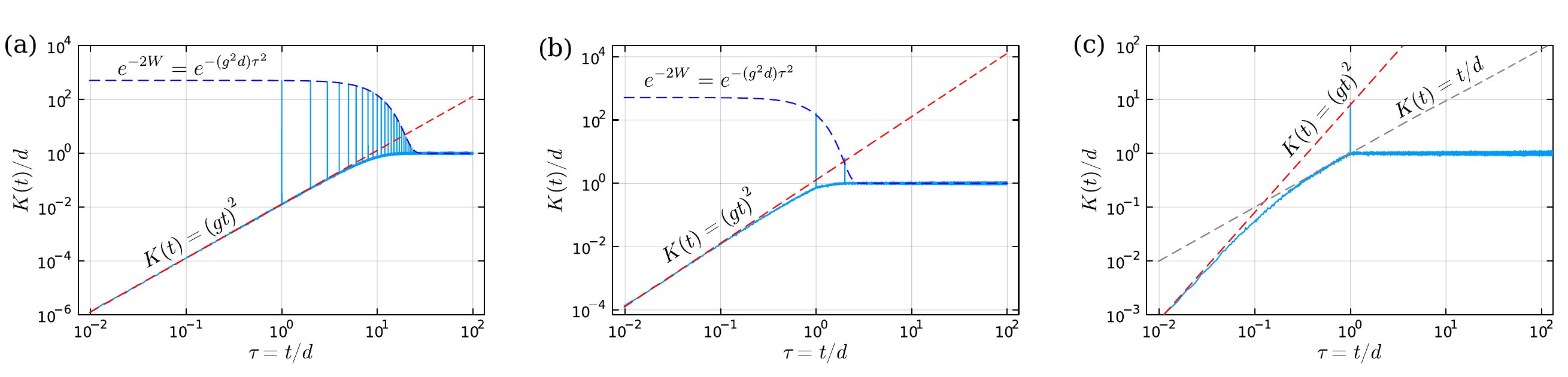}
    \caption{The SFF of the interpolating model $U = e^{-i g H} S$ for $g = 1/200$ (a), $g = 1/20$ (b), and $g = 1/8$ (c). We set $d = 512$ and average over 1000 samples. Dashed lines denote analytical fits implied by Eq.~\eqref{eq:SFF-single} or circular unitary ensemble universality. Permutation $S$ contains a single random cycle of length $d$.}
    \label{fig:SFF-IM}
\end{figure}

Finally, we take into account that in this approximation, all correlation functions break down into products of two-point correlators, which, in turn, decouple due to identity~\eqref{eq:two-point}. This allows us to estimate the SFF:
\beq \label{eq:SFF-single}
\begin{aligned}
    K(t) &\approx \sum_{j,k} e^{2 \pi i (k - j) t / d} \left\langle e^{i t (x_k - x_j)} \right\rangle 
    = \sum_{j,k} e^{2 \pi i (k - j) t / d} \left\langle e^{i t x_k} \right\rangle \left\langle e^{-i t x_j} \right\rangle \\
    &= d + \sum_{j \neq k} e^{2 \pi i (k - j) t / d} \exp\!\left( - \frac{t^2}{2} \la x_k x_k \ra - \frac{t^2}{2} \la x_j x_j \ra \right) \\
    &= d - d \exp\!\left( - \frac{g^2 t^2}{d} \right) + \left[ \frac{\sin(\pi t)}{\sin(\pi t/d)} \right]^2 \exp\!\left( - \frac{g^2 t^2}{d} \right).
\end{aligned} \eeq
So, similarly to the circular beta ensemble, toy interpolating model $U = e^{- i g H} S$ implies periodic SFF with period $t_H = d$ and Debye-Waller suppression $e^{-2W} = e^{-(g^2 d) \tau^2}$:
\beq \label{eq:hop-SFF}
K(t_H \tau) \approx d + e^{-(g^2 d) \tau^2} \left( d^2 - d \right),  \eeq
where $\tau$ is integer. We remind that this approximation is meaningful only when $g^2 d \ll 1$. Otherwise, it works only until the Thouless time $t_\mathrm{Th} \sim \sqrt{d} / g < t_H$. At larger times, the SFF approaches the standard ``ramp---plateau'' form inherent in the circular unitary ensemble, see Fig.~\ref{fig:SFF-IM}.

Finally, we extend these results to $N$ cycles of length $\{ d_1, \ldots, d_N \}$. If the spectrum of $S$ is not degenerate, all previous arguments work. Hence, similarly to the one-cycle case, we estimate eigenvalue correlations and obtain the same expression for the two-point function~\eqref{eq:two-point} and the SFF~\eqref{eq:SFF-single}:
\beq \label{eq:SFF-many}
K(t) \approx d - d \exp\!\left( - \frac{g^2 t^2}{d} \right) + \left[ \sum_{n=1}^N \frac{\sin(\pi t)}{\sin(\pi t/d_n)} \right]^2 \exp\!\left( - \frac{g^2 t^2}{d} \right). \eeq
If we remove random phase rotations from $S$, i.e., set all $\phi_k = 0$, the spectrum of $S$ becomes degenerate, since all sub-cycles have a common eigenvalue $E = 0$ (and other common eigenvalues if lengths $d_1, \ldots, d_n$ are not mutually prime). Corrections to such eigenvalues cannot be derived using the non-degenerate perturbation theory, but we can still approximately describe eigenvalue correlations by a theory of non-interacting particles. Assume that $\bar{x}_n$ are eigenvalues of the correlation matrix $C^{-1}_{mn} = \la x_m x_n \ra$ and introduce the transformation $x_m = \sum_n X_{mn} \bar{x}_n$. Then, the SFF is estimated as follows:
\beq \begin{aligned}
    K(t) &\approx d + \sum_{m \neq n} e^{2 \pi i t \left[ m / d(m) - n / d(n) \right]} \int \prod \bar{x}_k \exp\left[ i t \sum_p \left( X_{mp} - X_{np} \right) \bar{x}_p - \frac{1}{2} \sum_p C_p \bar{x}_p^2 \right] \\
    &= d + \sum_{m \neq n} e^{2 \pi i t \left[ m / d(m) - n / d(n) \right]} \exp\left[ - \frac{t^2}{4} \sum_k \frac{(X_{mk} - X_{nk})^2}{C_k} \right],
\end{aligned} \eeq
where function $d(m)$ returns the length of the cycle seeded by $m$ and $C_k$ are eigenvalues of $C_{mn}$. If number of degenerate eigenvalues is small, then the off-diagonal elements of the correlation matrix $C^{-1}_{mn}$ exponentially decay, so $X_{mk} \approx \delta_{mk}$ and $C_k \approx g^2 / d$ for all $k$. Hence, in this case, non-degenerate approximation works well. 
If number of degenerate eigenvalues grows large, non-degenerate approximation still qualitatively captures the behavior of the SFF, but fails to describe its fine structure. 
Note, however, that this situation is unlikely if we draw $S$ randomly, since in this case, it usually has one large cycle $d_\mathrm{max} \sim d$ that dominates the SFF expansion~\eqref{eq:SFF-many} and implies that at least $\mO(d)$ eigenvalues are non-degenerate.

\section{Local shallow circuit for the toy interpolating model}
\label{sec:Circuit}

\begin{figure}[t]
    \centering
    \includegraphics[width=0.5\linewidth]{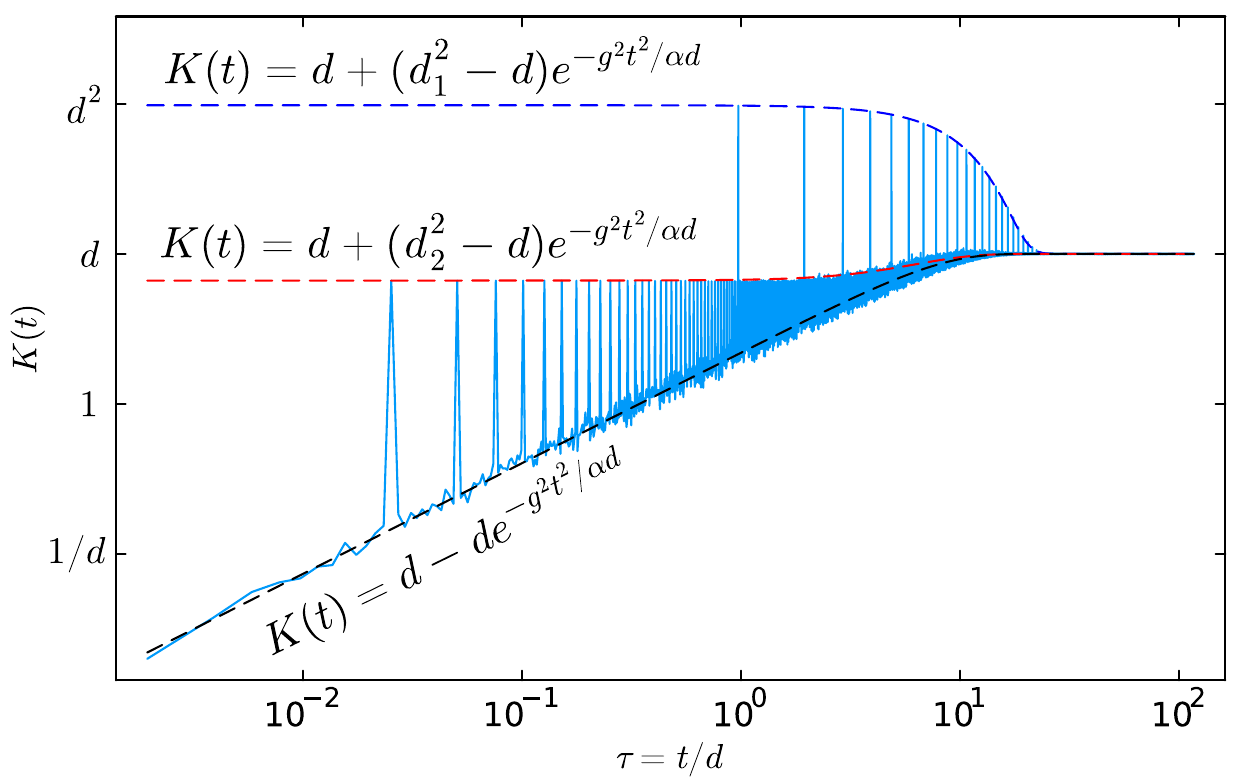}
    \caption{The SFF of the local interpolating model for $L = 9$ and $g = 1/500$. The permutation contains two cycles of length $d_1 = 499$ and $d_2 = 13$. The numerical factor $\alpha \approx 8$. We average over $10^4$ samples. Dashed lines denote analytical fits.}
    \label{fig:SFF-local}
\end{figure}

Although toy interpolating model $U = e^{-i g H} S$ is convenient for theoretical purposes, it is generally hard to work with matrices $H$ and $S$ defined on the entire Hilbert space. So, we suggest an alternative version of this model, in which both $H$ and $S$ are converted to depth-$n$ brick-work circuits acting on the chain of $L$ qubits (so, $d = 2^L$):
\beq \label{eq:Interpolation-local}
U = \prod_{k = 0}^{n-1} \prod_{m = 0}^{\lfloor L/n - 1 \rfloor} \exp\!\left(-i g H_{n m + k}\right) S_{n m + k}. \eeq
Here, $H_{n m + k}$ and $ S_{n m + k}$ are $2^n \times 2^n$ matrices acting in the basis of $n$ qubits starting from the qubit $n m + k$. Similarly to the global model, $S_{n m+ k}$ are classical permutations of computational basis states supplemented with random phase rotations, and $H_{n m + k}$ are random Hermitian matrices that make evolution~\eqref{eq:Interpolation-local} inherently non-classical. We assume that local permutations $S$ are fixed but drawn randomly. Then, we average the SFF of model~\eqref{eq:Interpolation-local} over the ensemble of matrices $H$. If blocks of the circuit are large enough ($n \ge 3$), the length of the largest cycle is comparable to the dimension of the full Hilbert space, $d_\mathrm{max} = \mathcal{O}(2^L)$. Hence, level statistics of this local model qualitatively reproduces the global model $U = e^{-i g H} S$, see Fig.~\ref{fig:SFF-local}.

